\documentstyle[12pt]{article}
     \def\lsim{\raise0.3ex\hbox{$<$\kern-0.75em\raise-1.1ex\hbox{$\sim$}}}
\def\gsim{\raise0.3ex\hbox{$>$\kern-0.75em\raise-1.1ex\hbox{$\sim$}}}
\def\noi{\noindent}
\def\nn{\nonumber}
\def\bea{\begin{eqnarray}}  \def\eea{\end{eqnarray}}
\def\beq{\begin{equation}}   \def\eeq{\end{equation}}

\def\beeq{\begin{eqnarray}} \def\eeeq{\end{eqnarray}}
\def\R{ {\rm R \kern -.31cm I \kern .15cm}}
\def\C{ {\rm C \kern -.15cm \vrule width.5pt \kern .12cm}}
\def\Z{ {\rm Z \kern -.27cm \angle \kern .02cm}}
\def\N{ {\rm N \kern -.26cm \vrule width.4pt \kern .10cm}}
\def\1{{\rm 1\mskip-4.5mu l} }

\def\lsim{\raise0.3ex\hbox{$<$\kern-0.75em\raise-1.1ex\hbox{$\sim$}}}
\def\gsim{\raise0.3ex\hbox{$>$\kern-0.75em\raise-1.1ex\hbox{$\sim$}}}
\def\noi{\noindent}
\def\nn{\nonumber}
\def\bea{\begin{eqnarray}} 
 \def\eea{\end{eqnarray}}
\def\beq{\begin{equation}}   
\def\eeq{\end{equation}}

\begin{document}
\begin{center}
{ \Large\bf Sum rules for leading and subleading form}\\
{\Large \bf  factors in
Heavy Quark Effective Theory}
{\Large\bf using the non-forward amplitude}
\vskip 5 truemm

{\bf F. Jugeau}\\
{\it Instituto de F\'isica Corpuscular, Valencia, Spain}
\vskip 5 truemm

{\bf A. Le Yaouanc, L. Oliver\footnote{Speaker at the conference} and J.-C. Raynal}\\
{\it Laboratoire de Physique Th\'eorique (UMR 8627 CNRS)}\\
{\it  Universit\'e de Paris Sud-XI, B\^atiment 210, 91405 Orsay cedex, France}
\end{center}

\begin{abstract}
 Within the OPE, we formulate new sum rules in Heavy Quark Effective Theory in the heavy quark limit and at order $1/m_Q$, using the non-forward amplitude. In the heavy quark limit, these sum rules imply that the elastic Isgur-Wise function $\xi
(w)$ is an alternate series in powers of $(w-1)$. Moreover, one gets
that the $n$-th derivative of $\xi (w)$ at $ w=1$ can be bounded by the
$(n-1)$-th one, and the absolute lower bound for the $n$-th derivative
$(-1)^n \xi^{(n)}(1) \geq {(2n+1)!! \over 2^{2n}}$.  Moreover, for the
curvature we find $\xi ''(1) \geq {1 \over 5} [4 \rho^2 + 3(\rho^2)^2]$
where $\rho^2 = - \xi '(1)$. These results
are consistent with the dispersive bounds, and they strongly reduce the
allowed region of the latter for $\xi (w)$. The method is extended to
the subleading quantities in $1/m_Q$. Concerning the perturbations of the Current, we derive new simple relations between the {\it functions} $\xi_3(w)$
and $\overline{\Lambda}\xi (w)$ and the sums $\sum\limits_n \Delta
E_j^{(n)} \tau_j^{(n)}(1) \tau_j^{(n)}(w)$ ($j = {1\over 2}, {3 \over
2})$, that involve leading quantities, Isgur-Wise functions
$\tau_j^{(n)}(w)$ and level spacings $\Delta E_j^{(n)}$. Our
results follow because the non-forward amplitude depends on three variables $(w_i, w_f, w_{if})
= (v_i\cdot v', v_f \cdot v', v_i \cdot v_f)$, and we consider the zero recoil frontier $(w,1,w)$ where only a
finite number of $j^P$ states contribute $\left ( {1\over 2}^+, {3
\over 2}^+\right )$. We also obtain new sum rules involving the elastic subleading form
factors $\chi_i(w)$ $(i = 1,2, 3)$ at order $1/m_Q$ that originate from
the ${\cal L}_{kin}$ and ${\cal L}_{mag}$ perturbations of the
Lagrangian. To the sum rules contribute only the same intermediate states $\left (
j^P, J^P\right ) = \left ( {1 \over 2}^-, 1^-\right ) , \left ( {3\over
2}^-, 1^-\right )$ that enter in the $1/m_Q^2$ corrections of the axial
form factor $h_{A_1}(w)$ at zero recoil. This allows to obtain a lower
bound on $- \delta_{1/m^2}^{(A_1)}$ in terms of the $\chi_i(w)$ and the
shape of the elastic IW function $\xi (w)$. An important theoretical implication is that $\chi '_1(1)$, $\chi_2(1)$ and $\chi ' _3(1)$ ($\chi_1(1) = \chi_3(1) = 0$ from Luke theorem)  must vanish when the slope and the curvature attain their
lowest values $\rho^2 \to {3 \over 4}$, $\sigma^2 \to {15 \over 16}$. These constraints should be taken into account in the exclusive determination of $|V_{cb}|$. 
\end{abstract}

\noi LPT Orsay 05-74  (October 2005)\\

\noi {\it International Workshop on Quantum Chromodynamics :
Theory and Experiment, Conversano (Bari, Italy), 16-20 June 2005 }
\newpage
\baselineskip=20 pt

We will expose the main results that we have obtained in Heavy Quark Effect Theory using the non-forward amplitude and the Operator Product Expansion. We will first examine results in the heavy quark limit on the shape of the Isgur-Wise function. The method is then generalized to the study of the $1/ m_Q$ perturbations, that are of two types, namely perturbations of the Current, and perturbations of the Lagrangian. This is a somewhat longer version of the talk than the one that will appear in the Proceedings.

\section{Heavy quark limit}

In the leading order of the heavy quark expansion of QCD, Bjorken sum
rule (SR) \cite{1r} relates the slope of the elastic Isgur-Wise (IW)
function $\xi (w)$, to the IW functions of the transitions between the
ground state and the $j^P = {1 \over 2}^+$, ${3 \over 2}^+$ excited
states, $\tau_{1/2}^{(n)}(w)$, $\tau_{3/2}^{(n)}(w)$, at zero recoil $w
= 1$ ($n$ is a radial quantum number). This SR leads to the lower bound
$- \xi '(1) = \rho^2 \geq {1 \over 4}$. Recently, a new SR was
formulated by Uraltsev in the heavy quark limit \cite{2r} involving
also $\tau_{1/2}^{(n)}(1)$, $\tau_{3/2}^{(n)}(1)$, that implies,
combined with Bjorken SR, the much stronger lower bound $\rho^2 \geq {3
\over 4}$, a result that came as a big surprise. In ref. \cite{3r}, in
order to make a systematic study in the heavy quark limit of
QCD, we have developed a manifestly covariant formalism within the
Operator Product Expansion (OPE). We did recover Uraltsev SR plus a new
class of SR. Making a natural physical assumption, this new class of SR
implies the bound $\sigma^2 \geq {5 \over 4} \rho^2$ where $\sigma^2$ is
the curvature of the IW function. Using this formalism including the
whole tower of excited states $j^P$, we have recovered rigorously the
bound $\sigma^2 \geq {5 \over 4} \rho^2$ plus generalizations that
extend it to all the derivatives of the IW function $\xi (w)$ at zero
recoil, that is shown to be an alternate
series in powers of $(w-1)$.

Using the OPE and the trace formalism in the heavy quark limit,
different initial and final four-velocities $v_i$ and $v_f$, and heavy
quark currents, where $\Gamma_1$ and $\Gamma_2$ are arbitrary Dirac
matrices $J_1 = \bar{h}_{v'}^{(c)}\ \Gamma_1\
h_{v_i}^{(b)}$, $J_2 = \bar{h}_{v_f}^{(b)}\ \Gamma_2\
h_{v'}^{(c)}$, the following sum rule can be written \cite{4r}~: 
\bea \label{2e} 
&&\Big \{ \sum_{D=P,V} \sum_n Tr \left [
\bar{\cal B}_f(v_f) \bar{\Gamma}_2 {\cal D}^{(n)}(v') \right ]
Tr \left [ \bar{\cal D}^{(n)}(v') \Gamma_1 {\cal B}_i(v_i)\right ]
\xi^{(n)} (w_i) \xi^{(n)} (w_f) \nn \\
&&+ \ \hbox{Other excited
states} \Big \}  = - 2 \xi(w_{if}) Tr \left [ \bar{\cal
B}_f(v_f) \bar{\Gamma}_2 P'_+ \Gamma_1 {\cal B}_i(v_i)\right ]\ .\eea

In this formula $v'$ is the intermediate meson four-velocity, $P'_+ =
{1 \over 2} (1 + {/ \hskip - 2 truemm v}')$ comes from the residue of
the positive energy part of the $c$-quark propagator, $\xi(w_{if})$ is
the elastic Isgur-Wise function that appears because one assumes $v_i
\not= v_f$. ${\cal B}_i$ and ${\cal B}_f$ are the $4 \times 4$ matrices
of the ground state $B$ or $B^*$ mesons and ${\cal D}^{(n)}$ those of
all possible ground state or excited state $D$ mesons coupled to $B_i$
and $B_f$ through the currents. In (\ref{2e}) we
have made explicit the $j = {1 \over 2}^-$ $D$ and $D^*$
mesons and their radial excitations of quantum number $n$. The explicit contribution of the other excited states is written below.\par

The variables $w_i$, $w_f$ and $w_{if}$ are defined as $w_i = v_i \cdot v'$, $w_f = v_f \cdot v'$, $w_{if} = v_i \cdot
v_f$. The domain of $(w_i$, $w_f$, $w_{if}$) is \cite{3r} 
\bea \label{4e} &&w_i, w_f \geq 1\nn\\
&&w_iw_f - \sqrt{(w_i^2 - 1)
(w_f^2 - 1)} \leq w_{if} \nn \\
&&\leq w_iw_f + \sqrt{(w_i^2 -1) ( w_f^2 - 1)} \
. \eea
The SR (\ref{2e}) writes $L\left ( w_i, w_f, w_{if} \right ) = R \left ( w_i,
w_f, w_{if} \right )$, where $L(w_i, w_f, w_{if})$ is the sum over the intermediate
charmed states and $R(w_i, w_f, w_{if})$ is the OPE side. Within the
domain (\ref{4e}) one can derive relatively to any of the variables
$w_i$, $w_f$ and $w_{if}$ and obtain different SR taking different limits to the frontiers
of the domain. \par

As in ref. \cite{3r} \cite{5newr}, we choose as initial and final states the $B$
meson ${\cal B}_i (v_i) = P_{i+} (- \gamma_5)$, ${\cal B}_f (v_f) = P_{f+} (- \gamma_5)$ and vector or axial currents projected along the $v_i$ and $v_f$
four-velocities
\beq
 \label{8Te}
  J_1 = \bar{h}_{v'}^{(c)}\ {/ \hskip - 2 truemm v}_i\ h_{v_i}^{(b)} \quad , \qquad J_2 = \bar{h}_{v_f}^{(b)}\  {/ \hskip - 2 truemm v}_f\  h_{v'}^{(c)}  
  \eeq
\noi we obtain SR (\ref{2e}) with the sum of all excited states $j^P$ in a
compact form~: 
\bea \label{9Se}
&&(w_i + 1) (w_f + 1) \sum_{\ell \geq
0} {\ell + 1 \over 2 \ell + 1} S_{\ell} (w_i, w_f, w_{if}) \sum_n \tau_{\ell + 1/2}^{(\ell)(n)}(w_i) \tau_{\ell + 1/2}^{(\ell
)(n)}(w_f)\nn\\
&& + \sum_{\ell \geq 1} S_{\ell} (w_i, w_f, w_{if})
\sum_n \tau_{\ell - 1/2}^{(\ell)(n)}(w_i) \tau_{\ell - 1/2}^{(\ell
)(n)}(w_f) = (1 + w_i+w_f+w_{if}) \xi(w_{if})\  \eea

We get, choosing instead the axial currents,
\beq \label{10Te} J_1 = \bar{h}_{v'}^{(c)}\ {/ \hskip - 2 truemm v}_i\
\gamma_5 \ h_{v_i}^{(b)} \ , \ J_2 = \bar{h}_{v_f}^{(b)}\ {/
\hskip - 2 truemm v}_f\ \gamma_5\ h_{v'}^{(c)} \ , \eeq
$$\sum_{\ell \geq 0} S_{\ell + 1}
(w_i, w_f, w_{if}) \sum_n \tau_{\ell + 1/2}^{(\ell)(n)}(w_i) \tau_{\ell
+ 1/2}^{(\ell )(n)}(w_f) + (w_i - 1) (w_f - 1)$$
\beq	
\label{11Re}
\sum_{\ell \geq
1} {\ell \over 2 \ell - 1} S_{\ell - 1} (w_i, w_f, w_{if}) \sum_n
\tau_{\ell - 1/2}^{(\ell)(n)}(w_i) \tau_{\ell - 1/2}^{(\ell )(n)}(w_f)
= - (1 - w_i-w_f+w_{if}) \xi (w_{if}) \  \eeq

Following the formulation of heavy-light states for arbitrary $j^P$
given by Falk \cite{4r}, we have defined in ref. \cite{3r} the IW
functions $\tau_{\ell + 1/2}^{(\ell)(n)}(w)$ and $\tau_{\ell -
1/2}^{(\ell)(n)}(w)$, $\ell$ and $j = \ell \pm {1
\over 2}$ being the orbital and total angular momentum
of the light cloud. \par

In (\ref{8Te}) and (\ref{10Te}) $S_n$ is given by 
\beq \label{12e} S_n = v_{i\nu_1} \cdots v_{i\nu_n}\ v_{f\mu_1} \cdots
v_{f\mu_n} \sum_{\lambda} \varepsilon'^{(\lambda )*\nu_1 \cdots \nu_n} \
\varepsilon'^{(\lambda )\mu_1 \cdots \mu_n} \ .\eeq
\noi One can show \cite{3r}~:
\beq \label{14e} S_n = \sum_{0 \leq k \leq {n \over 2}}
C_{n,k} (w_i^2 - 1)^k (w_f^2 - 1)^k (w_i w_f - w_{if})^{n-2k} \eeq
\noi with $C_{n,k} = (-1)^k {(n!)^2 \over (2n) !} \ {(2n - 2k) !
\over k! (n-k) ! (n-2k)!}$.\par 

\noi From the sum of (\ref{9Se}) and (\ref{11Re}) one obtains, differentiating relatively to
$w_{if}$ \cite{5newr} $(\ell \geq 0)$~:
\beq \label{16e} 
\xi^{(\ell
)} (1) = {1 \over 4} \ (-1)^{\ell} \ \ell ! \left \{ {\ell + 1 \over 2
\ell + 1} 4 \sum_n \left [ \tau_{\ell + 1/2}^{(\ell )(n)}(1) \right
]^2 + \sum_n \left [ \tau_{\ell - 1/2}^{(\ell -1)(n)}(1) \right
]^2 + \sum_n \left [ \tau_{\ell - 1/2}^{(\ell)(n)}(1) \right ]^2 \right
\}   \ .\eeq
\noi This relation shows that $\xi (w)$ is an alternate series in
powers of $(w-1)$. Equation (\ref{16e}) reduces to Bjorken SR \cite{1r}
for $\ell = 1$. Differentiating (\ref{11Re}) relatively to $w_{if}$ and
making $w_i=w_f = w_{if} = 1$ one obtains~: 
\beq \label{17He} \xi^{(\ell )} (1) = \ell !\
(-1)^{\ell} \sum_n \left [ \tau_{\ell + 1/2}^{(\ell)(n)}(1) \right ]^2
\quad (\ell \geq 0) \ . \eeq
\noi Combining (\ref{16e}) and (\ref{17He}) one obtains a SR for all $\ell$ that reduces to Uraltsev SR
\cite{2r} for $\ell = 1$. From (\ref{16e}) and (\ref{17He}) one obtains~:
 \beq \label{19e}
 (-1)^{\ell} \ \xi^{(\ell)} (1) = {1 \over 4} \ {2 \ell
+ 1 \over \ell} \ell ! \left \{ \sum_n \left [ \tau_{\ell - 1/2}^{(\ell
- 1)(n)}(1) \right ]^2 + \sum_n \left [ \tau_{\ell -
1/2}^{(\ell)(n)}(1) \right ]^2 \right \} \ . \eeq
\noi implying
$$(-1)^{\ell} \xi^{(\ell)}(1) \geq  {2\ell + 1 \over 4}
\left [  (-1)^{\ell - 1} \xi^{(\ell - 1)}(1) \right ] $$
\beq\label{20e} \geq  {(2\ell + 1)!! 
\over 2^{2\ell}} \eeq
\noi that gives, in particular, for the lower cases,
\beq
\label{22Qe}
- \xi ' (1) = \rho^2 \geq {3 \over 4} \quad , \quad \xi '' (1) \geq {15 \over 16}
\eeq

Considering systematically the derivatives of the SR (\ref{9Se}) and (\ref{11Re}) relatively to $w_i$, $w_f$, $w_{if}$ with the boundary conditions $w_{if} 
= w_i = w_f = 1$, one obtains a new SR:
\beq
\label{25e}
{4 \over 3} \rho^2 + (\rho^2)^2 - {5 \over 3} \sigma^2 + \sum_{n\not= 
0} |\xi^{(n)'}(1)|^2 = 0
\eeq
\noi that implies~:
  \beq
\label{27Qe}
\sigma^2 \geq {1 \over 5} \left [ 4 \rho^2 + 3(\rho^2)^2 \right ] \ .
\eeq
There is a simple intuitive argument to understand the term ${3 \over 
5} (\rho^2)^2$ in the best bound (\ref{27Qe}), namely the 
non-relativistic quark
model, i.e. a non-relativistic light quark $q$ interacting with a 
heavy quark $Q$ through a potential. The form factor has the simple form~:
\beq
\label{28Le}
F({\bf k}^2)= \int d {\bf r} \ \varphi^+_0(r) \exp \left ( i {m_q 
\over m_q + m_Q} {\bf k} \cdot {\bf r} \right ) \varphi_0 (r)
\eeq
\noi where $\varphi_0(r)$ is the ground state radial wave function. Identifying the non-relativistic IW function $\xi_{NR}(w)$ with 
the form factor $F({\bf k}^2)$ (\ref{28Le}),  one can prove that,
\beq
\label{31Ze}
\sigma_{NR}^2 \geq {3 \over 5} \ \left  [ \rho^2_{NR} \right ]^2 \ .
\eeq

Thus, the non-relativistic limit is a good guide-line to study the
shape of the IW function $\xi (w)$. We have recently generalized the
bound (\ref{31Ze}) to all the derivatives of $\xi_{NR}(w)$. The method uses
the positivity of matrices of moments of the ground state wave function
\cite{6newr}. We have shown that the method can be generalized to the 
function $\xi(w)$ of QCD.\par

An interesting phenomenological remark is that the simple 
parametrization for the IW function \cite{5r}
\beq
\label{32e}
\xi (w) = \left ( {2 \over w + 1} \right )^{2 \rho^2}
\eeq
\noi satisfies the inequalities (\ref{20e}), (\ref{27Qe}) if 
$\rho^2 \geq {3 \over 4}$. 
\par

The result (\ref{20e}), that shows that all derivatives at zero recoil
are large, should have important phenomenological implications for the
empirical fit needed for the extraction of $|V_{cb}|$ in $B \to D^*\ell
\nu$.  The usual fits to extract $|V_{cb}|$
using a linear or linear plus quadratic dependence of $\xi (w)$ are not
accurate enough.\par

A considerable effort has been developed to formulate dispersive 
constraints on the shape of the form factors in $\bar{B} \to D^*\ell 
\nu$ \cite{7r}-\cite{8r}, at finite mass. \par

Our approach, based on Bjorken-like 
SR, holds {\it in the physical region} of the semileptonic decays 
$\bar{B} \to D^{(*)}\ell \nu$ and
{\it in the heavy quark limit}. The two approaches are quite different in spirit and in their 
results. \par

Let us consider the main results of ref. \cite{8r} 
summarized by the one-parameter formula
\beq
\label{35e}
\xi(w)  \cong 1 - 8 \rho^2 z + (51\rho^2 - 10)z^2 - 
(252 \rho^2 - 84) z^3
\eeq
\noi with the variable $z(w)$ defined by
\beq
\label{36e}
z = {\sqrt{w+1} - \sqrt{2}\over \sqrt{w+1} + \sqrt{2}}
\eeq
\noi and the allowed range for $\rho^2$ being $- 0.17 < \rho^2 < 1.51$. This domain is considerably tightened by the lower bound on 
$\rho^2$~: ${3 \over 4} \leq \rho^2 < 1.51$, that shows that our type of bounds are complementary to the
bounds obtained from dispersive methods.

\section{${\bf 1/m_Q}$ perturbations of the Current }

In this Section, we follow the main lines of our paper \cite{9r}.
Our
starting point is the $T$-product
\beq \label{6a} T_{fi}(q) = i \int d^4x\ e^{-iq\cdot x} \
<B(p_f)|T[J_f(0)J_i(x) ]| B(p_i)> \eeq

\noi where $J_f(x)$, $J_i(y)$ are the currents (the convenient notation
for the subindices $i$, $f$ will appear clear below)~:
\beq \label{7a} J_f(x) = \overline{b}(x) \Gamma_f c(x) \qquad \qquad
J_i(y) = \overline{c}(y)\Gamma_i b(y) \eeq

\noi and $p_i$ is in general different from $p_f$. \par

Inserting in this expression hadronic intermediate states, $x^0 < 0$
receives contributions from the direct channel with hadrons with a
single heavy quark $c$, while $x^0 > 0$ receives contributions from
hadrons with $b\overline{c}b$ quarks, the $Z$ diagrams~: \bea
\label{8a} &&T_{fi}(q) =  \sum_{X_c} (2 \pi)^3
\ \delta^3( {\bf q} + {\bf p}_i - {\bf p}_{X_c}){<B_f|J_f(0)|X_c> <X_c|J_i(0)|B_i> \over q^0 + E_i - E_{X_c} + i \varepsilon}\nn \\ 
&&-\sum_{X_{\overline{c}bb}} (2\pi)^3 \ \delta^3 ( {\bf q} - {\bf p}_f
+ {\bf p}_{X_{\overline{c}bb}}) {<B_f|J_i(0)|X_{\overline{c}bb}>
<X_{\overline{c}bb}|J_f(0)|B_i> \over q^0 - E_f + E_{X_{\overline{c}bb}}- i \varepsilon}
\ . \eea

We consider the limit $m_c \gg m_b \gg \Lambda_{QCD}$. The difference between the two energy denominators is large
$q^0 - E_f + E_{X_{\overline{c}bb}} - \left ( q^0 + E_i - E_{X_c} 
\right ) \sim 2m_c$ and therefore, we can in this limit neglect the second term, and we
consider the imaginary part of the direct diagram, the first term
in (\ref{8a}), the piece proportional to $\delta \left ( q^0 + E_i - E_{X_c}\right )$. Our conditions are, in short, as follows~:
\beq
\label{no6bis}
\Lambda_{QCD} \ll m_b \sim m_c - q^0 \ll q^0 \sim m_c \ .
\eeq

\noi To summarize, we are considering 
the heavy quark limit for the $c$
quark, but we allow for a large finite mass for the $b$ quark. \par

In the conditions (\ref{no6bis}), or choosing the suitable integration
contour, we can write, integrating over
$q^0$
\beq
\label{no7}
T^{abs}_{fi}({\bf q}) \cong \sum_{X_c} (2 \pi)^3 \ \delta^3\left 
({\bf q} + {\bf p}_i - {\bf q}_{X_c}\right ) <B_f|J_f(0)|X_c>\ 
<X_c|J_i(0)|B_i>\ .
\eeq

\noi Finally, integrating over ${\bf q}_{X_c}$ and defining $v' = {q 
+ p_i \over m_c}$ one gets
\beq
\label{no8}
T^{abs}_{fi} \cong \sum_{D_n} <B_f(v_f)|J_f(0)|D_n(v')>\ <D_n(v')|J_i(0)|B_i(v_i)>
\eeq

\noi where we have denoted by $D_n(v')$ the charmed intermediate states.\par

The $T$-product matrix element $T_{fi}(q)$ (\ref{6a}) is given, 
alternatively, in terms of quarks and gluons, by the expression
\beq
\label{no9}
T_{fi} (q) = - \int d^4 x \ e^{-iq\cdot x} \  < 
B(p_f)|\overline{b}(0)\Gamma_f S_c (0, x) \Gamma_i b(x) |B(p_i)>
\eeq

\noi where $S_c (0, x)$ is the $c$ quark propagator in the background
of the soft gluon field \cite{13r}. \par

Since we are considering the absorptive part in the $c$ heavy quark
limit of the direct graph in (\ref{8a}), this quantity can be then
identified with (\ref{no9}) where $S_c(x,0)$ is replaced by the
following expression \cite{14r}
\beq
\label{no10}
S_c(0, x) \to e^{im_cv'\cdot x} \ \Phi_{v'}[0, x] D_{v'}(x)
\eeq

\noi where $D_{v'}(x)$ is the {\it cut} free propagator of a heavy quark
\beq
\label{no11}
D_{v'}(x) = P'_+ \int {d^4k \over (2 \pi)^4} \delta (k\cdot v') 
e^{ik\cdot x} = P'_+ \int_{-\infty}^{\infty} {dt \over 2\pi} \delta^4 
(x- v't)
\eeq

\noi with the positive energy projector defined by $P'_+ = {1 \over 2} (1 + {/ \hskip - 2 truemm v}')$.

The eikonal phase $\Phi_{v'}[0, x]$ in (\ref{no10}) corresponds to 
the propagation of the $c$ quark from the point $x = v't$ to the 
point 0, that is given by
\beq
\label{no13}
\Phi_{v'} [0, v't] = P \exp \left ( - i \int_0^t ds\ v'\cdot A(v's)\right )\ .
\eeq

\noi This quantity takes care of the dynamics of the soft gluons in
HQET along the classical path $x = v't$. \par

We obtain
\bea
\label{no15}
&&T^{abs}_{fi}(q) = \int d^4 x \ e^{-i(q-m_c v')\cdot x} 
\int_{-\infty}^{\infty} {dt \over 2 \pi} \delta^{4} (x - v't)\nn \\ 
&&<B(p_f) | \overline{b}(0) \Gamma_f
P'_+ \Phi_{v'} [0, x] \Gamma_i b(x) |B(p_i)>
 + \ O(1/m_c)  \ .
\eea
\noi Integrating over $x$ in (\ref{no15}) and making explicit (\ref{no13}),
\bea
\label{no16}
&&T^{abs}_{fi}(q) = \int_{-\infty}^{\infty} {dt \over 2\pi} \ 
e^{-i(q-m_c v')\cdot v't}  \\
&&<B(p_f) | \overline{b}(0) \Gamma_f
P'_+ P \exp\left ( - i \int_0^t ds\ v'\cdot A(v's)\right ) \Gamma_i 
b(v't) |B(p_i)>  + \ O(1/m_c)  \ .\nn 
\eea

Performing first the integration over $q^0$ one obtains simply $\delta
(v'^0t)$, and making the trivial integration over
$t$ one obtains finally the OPE matrix element~:
\beq
\label{no17}
T^{abs}_{fi} \cong \ <B(p_f) | \overline{b}(0) \Gamma_f {1 + 
{/\hskip -2 truemm v}' \over 2v'^0} \Gamma_i b(0)|B(p_i)> \ + \ 
O(1/m_c)  \ .
\eeq
Therefore, we end up with the sum rule
\bea
\label{24NR}
&&\sum_{D_n} <B_f(v_f)|J_f(0)|D_n(v')>\ <D_n(v')|J_i(0)|B_i(v_i)>\nn \\
&&=\ <B(v_f) |\overline{b}(0)\Gamma_f {1 + {/\hskip -2 truemm v}' 
\over 2v'^0} \Gamma_i b(0)|B(v_i)>\ + \ O(1/m_c)
\eea
\noi that is valid for {\it all powers} of an expansion in $1/m_b$, 
but only to leading order in $1/m_c$. \par

On the other hand, making use of the HQET equations of motion, the
field $b(x)$ in (\ref{24NR}) can be decomposed into upper and lower
components as follows \cite{15r}
\beq
\label{no14}
b(x) = e^{-im_bv\cdot x} \left ( 1 + {1 \over 2m_b + iv\cdot 
\overrightarrow{D}} \ i\overrightarrow{/\hskip - 3 truemm D}\right ) 
h_{v}(x)
\eeq

\noi where the second term corresponds to the lower components and 
can be expanded in a series in powers of $D_{\mu}/m_b$, and $v$ is an 
arbitrary four-velocity.\par

Keeping the first order in $1/m_b$, the sum rule reads
\bea
\label{no18}
&&\sum_{D_n} <B_f(v_f)|J_f(0)|D_n(v')>\ <D_n (v')|J_i(0)|B_i(v_i)>\nn \\
&&= \ <B(p_f) |\overline{h}_{v_f}(0)\Gamma_f{1 + {/\hskip -2 truemm 
v}' \over 2v'^0} \Gamma_i h_{v_i}(0)|B(p_i)> \ \nn \\
&&+{1 \over 2 m_b}<B(p_f)| \overline{h}_{v_f}(0) \Big [ 
(-i\overleftarrow{/\hskip - 3 truemm D}) \Gamma_f{1 + {/\hskip -2 
truemm v}' \over 2v'^0}\Gamma_i
+ \Gamma_f{1 + {/\hskip -2 truemm v}' \over 2v'^0}\Gamma_i 
(i\overrightarrow{/\hskip - 3 truemm D}) \Big ]h_{v_i}(0)|B(p_i)> \nn \\
&& + \ O(1/m_c) + O(1/m_b^2) \ .
\eea
Therefore, in the OPE side we have, besides the leading dimension 3
operator

\beq \label{18a} O^{(3)} = \overline{h}_{v_f} \Gamma_f P'_+ \Gamma_i
h_{v_i} \eeq

\noi the dimension 4 operator
\beq \label{19a} O^{(4)} = \overline{h}_{v_f} \left [ (- i
\overleftarrow{/\hskip - 3 truemm D}) \Gamma_f P'_+ \Gamma_i + \Gamma_f
P'_+ \Gamma_i (i \overrightarrow{/\hskip - 3 truemm D})\right ] 
h_{v_i} \ . \eeq

In the SR we have to compute the l.h.s. including terms of order
$1/2m_b$. These terms have been parametrized by Falk and Neubert \cite{17r} for
the ${1 \over 2}^-$ doublet and by Leibovich et al. \cite {16r} for the transitions
between the ground state ${1 \over 2}^-$ and the ${1 \over 2}^+$, ${3
\over 2}^+$ excited states.  \par

A remark is in order here, that was already made in ref. \cite{11r}.
Had we taken higher moments of the form $\int dq^0(q^0)^n
T_{fi}^{abs}(q^0)$ ($n > 0$), instead of the lowest one $n = 0$, the
integration over $q^0$ that leads to the simple sum rules
(\ref{24NR}) or (\ref{no18}) would involve higher dimension operators, giving
a whole tower of sum rules \cite{18r,14r}, even in the {\it
leading} heavy quark limit. Our point of view in this paper is
different. We consider the lowest moment $n = 0$, while we expand in
powers of $1/m_b$, keeping the first order in this parameter.\par

Concerning the OPE side in (\ref{no18}), the dimension 4 operator 
$O^{(4)}$ (\ref{19a}) is nothing
else but the $1/m_b$ perturbation of the heavy current $O^{(3)} =
\overline{h}_{v_f}\Gamma_fP'_+\Gamma_i h_{v_i}$ since this operator,
containing the Dirac matrix $\Gamma_f P'_+ \Gamma_i$ between heavy 
quark fields, can be considered
as a heavy quark current. Indeed, following Falk and Neubert, the 
$1/m_b$ perturbation of any
heavy quark current $\overline{h}_{vf}\Gamma h_{vi}$ is given by
\beq \label{20a} \overline{h}_{v_f}\left ( -{i \overleftarrow{/\hskip -
3 truemm D} \over 2m_b}\right ) \Gamma h_{v_i} + \overline{h}_{v_f}\Gamma\left
( {i \overrightarrow{/\hskip - 3 truemm D} \over 2m_b}\right )
h_{v_i} \ . \eeq

However, this perturbation of the current does not exhaust all
perturbations in $1/m_b$. We need also to compute the
perturbation of the initial and final wave functions $|B_i(v_i)>$,
$|B_f(v_f)>$ due to the kinetic and magnetic perturbations of the
Lagrangian. This can be done easily following also the prescriptions of
Falk and Neubert to compute these corrections in $1/m_b$ for the leading
matrix element $<B_f(v_f)|\overline{h}_{vf}\Gamma_f P'_+ \Gamma_i
h_{vi}|B_i(v_i)>$.

The final result is the following. The subleading quantities, functions of $w$,
$\overline{\Lambda}\xi (w)$ and $\xi_3(w)$ \cite{17r} can be expressed in terms of leading
quantities, namely the ${1\over 2}^- \to {1 \over 2}^+, {3 \over 2}^+$
IW functions $\tau_j^{(n)}(w)$ and the corresponding level spacings
$\Delta E_j^{(n)}$ $(j = {1 \over 2}, {3 \over 2})$ \cite{9r}
\beq \label{44Ye} \overline{\Lambda} \xi (w) = 2(w+1) \sum_n \Delta
E_{3/2}^{(n)} \ \tau_{3/2}^{(n)}(1) \ \tau_{3/2}^{(n)}(w)+ \ 2
\sum_n \Delta E_{1/2}^{(n)} \ \tau_{1/2}^{(n)}(1) \ \tau_{1/2}^{(n)}(w)
\eeq
\beq \label{45Ye} \xi_3 (w) = (w+1) \sum_n \Delta E_{3/2}^{(n)} \
\tau_{3/2}^{(n)}(1) \ \tau_{3/2}^{(n)}(w)- \ 2 \sum_n \Delta
E_{1/2}^{(n)} \ \tau_{1/2}^{(n)}(1) \ \tau_{1/2}^{(n)}(w) \ . \eeq

\noi These quantities reduce to known SR for $w=1$, respectively
Voloshin SR \cite{10r} and a SR for $\xi_3(1)$ \cite{11r,2r}, and
generalizes them for all $w$.\par

The comparison of (\ref{44Ye}), (\ref{45Ye}) with the results of the BT quark model
\cite{5r} is very encouraging. Within this scheme $\xi (w)$ is given by
(\ref{32e}) with $\rho^2 = 1.02$, while one gets, for the $n = 0$ states
\beq \label{65} \tau_{j}^{(0)}(w) = \tau_{j}^{(0)}(1)\left ( {2
\over w+1}\right )^{2\sigma_{j}^2} 
\eeq

\noi with $\tau_{3/2}^{(0)}(1) = 0.54$, $\sigma_{3/2}^2 = 1.50$,
$\tau_{1/2}^{(0)}(1) = 0.22$ and $\sigma_{1/2}^2 = 0.83$. Assuming the
reasonable saturation of the SR with the lowest $n = 0$ states
\cite{5r}, one gets, from the first relation (\ref{44Ye}), a sensibly
constant value for $\overline{\Lambda} = 0.513 \pm 0.015$. \par

\section{${\bf 1/m_Q}$ perturbations of the Lagrangian}

We follow closely our recent work \cite{19r}. Instead of using the OPE, we will simply use the definition of the
subleading elastic ${1\over 2}^- \to {1 \over 2}^-$ functions
$\chi_i(w)$ $(i = 1,2,3)$ \cite{17r}
\bea
\label{5e}
&&<D(v') |i \int dxT [J^{cb} (0), {\cal L}_v^{(b)}(x)]|B(v)>\ = 
{1 \over 2m_b} \Big \{ - 2 \chi_1(w) Tr \left [\overline{D}(v') \Gamma B(v)\right ] \nn\\
&&+ {1 \over 2} Tr\left  [ A_{\alpha\beta}(v,v') \overline{D}(v') \Gamma P_+ i \sigma^{\alpha\beta} B(v)\right ] \Big \}\\
&&\nn\\
&&<D(v') |i \int dxT [J^{cb} (0), {\cal L}_{v'}^{(c)}(x)]|B(v)>\ = 
{1 \over 2m_c} \Big \{ - 2 \chi_1(w) Tr \left [\overline{D}(v') \overline{\Gamma} B(v)\right ] \nn\\
&&- {1 \over 2} Tr\left  [ \overline{A}_{\alpha\beta}(v',v) \overline{D}(v') i \sigma^{\alpha\beta} P'_+ \overline{\Gamma} B(v)\right ] \Big \}
\label{6e}
\eea
\noi with
\bea
\label{7e}
&&A_{\alpha \beta}(v,v') = - 2 \chi_2 (w) \left ( v'_{\alpha} \gamma_{\beta} - v'_{\beta} \gamma_{\alpha}\right ) + 4 \chi_3 (w) i \sigma_{\alpha \beta} \nn \\
&&\overline{A}_{\alpha \beta}(v',v) = - 2 \chi_2 (w) \left ( v_{\alpha} \gamma_{\beta} - v_{\beta} \gamma_{\alpha}\right ) - 4 \chi_3 (w) i \sigma_{\alpha \beta} 
\eea

\noi where $\overline{A} = \gamma^0 A^+\gamma^0$ denotes the Dirac
conjugate matrix, the current $J^{cb}(0)$ denotes
\beq
\label{8e}
J^{cb} = \overline{h}_{v'}^{(c)} \Gamma h_v^{(b)}
\eeq

\noi where $\Gamma$ is any Dirac matrix, and ${\cal L}_v^{(Q)}(x)$ is given by 
\beq
\label{9e}
{\cal L}_v^{(Q)} = {1 \over 2 m_Q} \left [ O_{kin, v}^{(Q)} + O_{mag, v}^{(Q)}\right ]
\eeq

\noi with
\beq
\label{10e}
O_{kin, v}^{(Q)} = \overline{h}_v^{(Q)}(iD)^2 h_v^{(Q)} \qquad O_{mag, v}^{(Q)} = {g_s \over 2} \overline{h}_v^{(Q)} \sigma_{\alpha \beta} G^{\alpha\beta} h_v^{(Q)} \ .
\eeq

In relations (\ref{5e})-(\ref{7e}), the $\chi_i(w)$ $(i= 1, 2, 3)$ have dimensions of mass, and correspond to the definition given by Luke \cite{20r}.

We will now insert intermediate states in the $T$-products (\ref{5e}).
We can separately consider ${\cal L}_{kin}^{(b)}$ or ${\cal
L}_{mag}^{(b)}$. The possible $Z$-diagrams involving heavy quarks
contributing to the $T$-products are suppressed by the heavy quark mass
since they are $b\overline{c}c$ intermediate states.\par

Conveniently choosing the initial and final states, we find the
following results~:\par

(1) With ${\cal L}_{kin,v}^{(b)}$, pseudoscalar initial state $B(v) = P_+
(- \gamma_5)$ and pseudoscalar final
state $\overline{D}(v') = \gamma_5P'_+$, one finds
\beq
\label{13e}
2 \chi_1(w) = \sum_{n\not= 0} {1 \over \Delta E_{1/2}^{(n)}}\ \xi^{(n)}(w) {<B^{(n)}(v)|O_{kin, v}^{(b)}(0)|B(v)> \over \sqrt{4m_{B^{(n)}} m_B}}\ .
\eeq

In the preceding expressions the energy denominators are $\Delta E_{1/2}^{(n)} = E_{1/2}^{(n)} - E_{1/2}^{(0)}$ $(n \not= 0)$.

(2) Consider ${\cal L}_{mag, v}^{(b)}$, pseudoscalar initial state $B(v) =
P_+ (- \gamma_5)$ and pseudoscalar final state $\overline{D}(v') =
\gamma_5 P'_+$. Because of parity conservation by the strong
interactions, the intermediate states $B^{(n)}$ must have the same
parity than the initial state $B$. Moreover, ${\cal L}_{mag, v}^{(b)}$ being a
scalar and producing transitions at zero recoil, the spin of $B$ and
$B^{(n)}$ must be the same. Therefore, only pseudoscalar intermediate
states $B^{(n)}(0^-)$ can contribute, only states with $j
= {1 \over 2}^-$. One finds, for any current (\ref{8e})
\beq
\label{17e}
- 4(w-1) \chi_2(w) + 12 \chi_3 (w) = \sum_{n\not= 0} {1 \over \Delta E_{1/2}^{(n)}} \xi^{(n)} (w)
 {<B^{(n)}(v) | O_{mag, v}^{(b)}(0)|B(v)> \over \sqrt{4m_{B^{(n)}} m_B}} \ . 
\eeq

\noi It is remarkable that this linear combination depends only on ${1
\over 2}^-$ intermediate states. \par

(3) Consider ${\cal L}_{mag, v}^{(b)}$ and a vector initial state $B^*(v ,
\varepsilon ) = P_+ {/ \hskip - 2 truemm \varepsilon}$ and pseudoscalar final state
$\overline{D}(v') = \gamma_5 P'_+$. Now we will have vector
$1^-$ intermediate states, either $B^{*(n)}\left ( {1 \over 2}^-,
1^-\right )$ or $B^{*(n)}\left ( {3 \over 2}^-, 1^-\right )$. For
the latter, we have to compute the current matrix element
\beq
\label{18e}
<D(v')|J^{cb}(0)|B^{*(n)}\left ( \textstyle{{3 \over 2}^-, 1^-}\right )(v, \varepsilon )>
= \tau_{3/2}^{(2)(n)}(w) Tr \left [ \overline{D}(v') \Gamma F_v^{\sigma} v'_{\sigma}\right ] 
\eeq
\noi where the $\left ( {3 \over 2}^-, 1^-\right )$ operator is given by
\beq
\label{19e}
F_v^{\sigma} = \sqrt{{3 \over 2}} P_+ \varepsilon_{\nu} \left [ g^{\sigma \nu} - {1 \over 3} \gamma^{\nu} \left ( \gamma^{\sigma} + v^{\sigma}\right ) \right ]
\eeq

\noi obtained from the $\left ( {3 \over 2}^+, 1^+\right )$
operator defined by Leibovich et al. (formula (2.5) of \cite{16r}),
multiplying by $(-\gamma_5)$ on the right \cite{4r}. The Isgur-Wise
functions $\tau_{3/2}^{(2)(n)}(w)$ correspond to ${1 \over 2}^- \to
{3 \over 2}^-$ transitions, the superindex ($\ell$) meaning the
orbital angular momentum \cite{5newr} \cite{4r} \cite{9r}. As noticed by
Leibovich et al. \cite{16r}, on general grounds the IW functions
$\tau_{3/2}^{(2)(n)}(w)$ do not vanish at zero recoil.\par

One finds, for any curent (\ref{8e}), finally 
\bea
\label{21e}
&&- 4 \chi_2 (w) (\varepsilon \cdot v') Tr \left [ \overline{D}(v') \Gamma P_+\right ] + 4 \chi_3 (w) Tr \left [ \overline{D}(v') \Gamma P_+ {/  \hskip - 1.5 truemm \varepsilon} \right ] \nn \\ 
&&= - Tr \left [ \overline{D}(v') \Gamma B^*(v, \varepsilon)\right ] \sum_{n\not= 0} {1 \over \Delta E_{1/2}^{(n)}} \xi^{(n)} (w)
{<B^{*(n)}(v, \varepsilon) | O_{mag, v}^{(b)}(0)|B^*(v, \varepsilon)> \over \sqrt{4m_{B^{*(n)}} m_{B^*}}} \nn \\
&&+ \left \{ \sqrt{{3 \over 2}} (\varepsilon \cdot v') Tr \left [ \overline{D}(v') \Gamma P_+\right ]  - {1 \over \sqrt{6}} (w-1) Tr \left [ \overline{D}(v') \Gamma P_+ {/  \hskip - 1.5 truemm \varepsilon}\right ]\right \} \nn \\
&&\sum_n {1 \over \Delta E_{3/2}^{(n)}} \tau_{3/2}^{(2)(n)} (w) {<B^{*(n)}_{3/2}(v, \varepsilon) | O_{mag, v}^{(b)}(0)|B^*(v, \varepsilon)> \over \sqrt{4m_{B^{*(n)}_{3/2}} m_{B^*}}}\ . 
\eea

The energy denominators are $\Delta E_{1/2}^{(n)} = E_{1/2}^{(n)} - E_{1/2}^{(0)}$ $(n \not= 0)$ and $\Delta E_{3/2}^{(n)} = E_{3/2}^{(n)} - E_{1/2}^{(0)}$ $(n \geq 0)$.

One can obtain other linearly independent relations, taking $\Gamma = \gamma_{\mu} \gamma_5$. 

Since the two four vectors $(v'_{\mu} - v_{\mu})$ and
$[(w-1)\varepsilon_{\mu} + (\varepsilon \cdot v' ) v_{\mu}  ]$ can be
chosen to be independent, one obtains independent sum rules for
$\chi_2(w)$ and $\chi_3(w)$.

To summarize, making explicit the $c$ flavor, we
have obtained the sum rules 
\beq
\label{26e}
\chi_1(w) = {1 \over 2} \sum_{n\not= 0} {1 \over \Delta E_{1/2}^{(n)}} \xi^{(n)} (w) {<D^{(n)}(v) | O_{kin, v}^{(c)}(0)|D(v)> \over \sqrt{4m_{D^{(n)}} m_{D}}} 
\eeq 
\beq
\label{27e}
\chi_2(w) = - {3 \over 4 \sqrt{6}} \sum_n {1 \over \Delta E_{3/2}^{(n)}} \tau_{3/2}^{(2)(n)} (w)
 {<D^{*(n)}_{3/2}(v, \varepsilon) | O_{mag, v}^{(c)}(0)|D^*(v, \varepsilon)> \over \sqrt{4m_{D^{*(n)}_{3/2}} m_{D^*}}} \eeq
\bea
\label{28e}
&&\chi_3(w) = - {1 \over 4} \sum_{n\not= 0} {1 \over \Delta E_{1/2}^{(n)}} \xi^{(n)} (w) {<D^{*(n)}(v, \varepsilon) | O_{mag, v}^{(c)}(0)|D^*(v, \varepsilon)> \over \sqrt{4m_{D^{*(n)}} m_{D^*}}} \nn \\
&&\qquad - \ {w-1 \over 4\sqrt{6}} \sum_{n} {1 \over \Delta E_{3/2}^{(n)}} \tau_{3/2}^{(2)(n)} (w) {<D^{*(n)}_{3/2}(v, \varepsilon) | O_{mag, v}^{(c)}(0)|D^*(v, \varepsilon)> \over \sqrt{4m_{D^{*(n)}_{3/2}} m_{D^*}}} 
\eea

There are a number of striking features in relations
(\ref{26e})-(\ref{28e}).\par

(i) One should notice that {\it elastic subleading form factors of the
Lagrangian type} are given in terms of {\it leading IW functions},
namely $\xi^{(n)}(w)$ and $\tau_{3/2}^{(2)(n)}(w)$, and {\it
subleading} form factors {\it at zero recoil}. \par

(ii) $\chi_1 (w)$ is given in terms of matrix elements of ${\cal
L}_{kin}$, as expected from the definitions (\ref{5e})-(\ref{6e}) and
involve transitions ${1\over 2}^- \to {1 \over 2}^-$. \par

(iii) The {\it elastic subleading magnetic form factors} $\chi_2(w)$ and
$\chi_3(w)$ involve $D^*(1^-) \to D^{*(n)} (1^-)$ transitions ${1 \over 2}^- \to {1 \over 2}^-$ and ${1 \over 2}^- \to {3 \over 2}^-$.\par

(iv) $\chi_1(w)$ and $\chi_3(w)$ satisfy, as they should, Luke theorem
\cite{20r},
\beq
\label{29e}
\chi_1(1) = \chi_3(1) = 0
\eeq

\noi because the ${1 \over 2}^- \to {1 \over 2}^-$ IW functions at zero recoil satisfy
\beq
\label{30e}
\xi^{(n)} (1) = \delta_{n,0}
\eeq

(v) There is a linear combination of $\chi_2(w)$ and $\chi_3(w)$ that
gets only contributions from ${1 \over 2}^- \to {1 \over 2}^-$
transitions, namely
\beq
\label{31e}
-4(w-1)\chi_2(w) + 12 \chi_3(w) = - 3 \sum_{n\not= 0} {1 \over \Delta E_{1/2}^{(n)}} \xi^{(n)} (w)
 {<D^{*(n)}(v, \varepsilon) | O_{mag, v}^{(c)}(0)|D^*(v, \varepsilon)> \over \sqrt{4m_{D^{*(n)}} m_{D^*}}}  
\eeq

\noi where the factor $-3$ is in consistency with (\ref{17e}), shifting
from vector to pseudoscalar mesons. \par

It is well-known that the determination of $|V_{cb}|$ from the $\overline{B} \to
D^*\ell \nu$ differential rate at zero recoil depends on the value of
$h_{A_1}(1)$. One interesting point is that precisely the subleading matrix elements
of $O_{kin}$ and $O_{mag}$ at zero recoil, that enter in the SR
(\ref{26e})-(\ref{28e}), are related to the quantity $|h_{A_1}(1)|$, as we will see now.

The following SR follows from the OPE \cite{18r} \cite{16r},
\bea
\label{33e}
&&|h_{A_1}(1)|^2 + \sum_n {|<D^{*(n)}\left ( {1 \over 2}^-, {3 \over 2}^-\right ) (v, \varepsilon )| \vec{A}|B(v)>|^2 \over 4m_{D^{*(n)}} m_B}\nn \\
&&= \eta_A^2 - {\mu_G^2 \over 3 m_c^2} - {\mu_{\pi}^2 - \mu_G^2 \over 4} \left ( {1 \over m_c^2} + {1 \over m_b^2} + {2 \over 3m_cm_b}\right )
\eea
\noi where $D^{*(n)}$ are $1^-$ excited states, and 
\bea
\label{34e}
&&\mu_{\pi}^2 = {1 \over 2m_B} \ <B(v)|\overline{h}_v^{(b)}(iD)^2 h_v^{(b)}|B(v)> \nn \\
&&\mu_{G}^2 = {1 \over 2m_B} \ <B(v)|\overline{h}_v^{(b)}{g_s \over 2} \sigma_{\alpha\beta}G^{\alpha\beta} h_v^{(b)}|B(v)> 
\eea
\noi In relation (\ref{33e}) one assumes the states at rest $v = (1,
{\bf 0})$ and the axial current is space-like, orthogonal to $v$. \par

In the l.h.s. of relation (\ref{33e}), 
\beq
\label{35e}
h_{A_1}(1) = \eta_{A_1} + \delta_{1/m^2}^{(A_1)}
\eeq

\noi ($\eta_{A_1} = 1 +$ radiative corrections) because there are no first order $1/m_Q$ corrections due to Luke theorem. The sum over the
squared matrix elements of $B \to D^{*(n)}(1^-)$ transitions contains
two types of possible contributions, corresponding to $D^{*(n)}\left (
{1 \over 2}^-, 1^-\right )$ $(n \not= 0)$, and $D^{*(n)}\left ( {3
\over 2}^-, 1^-\right )$ $(n \geq 0)$. The r.h.s. of (\ref{33e})
exhibits the OPE at the desired order. From the decomposition between
radiative corrections and $1/m_Q^2$ corrections (\ref{35e}) one gets,
from (\ref{33e}), neglecting higher order terms,
\bea
\label{36e}
&&- \delta_{1/m^2}^{(A_1)} = {\mu_G^2 \over 6 m_c^2} + {\mu_{\pi}^2 - \mu_G^2 \over 8} \left ( {1 \over m_c^2} + {1 \over m_b^2} + {2 \over 3m_cm_b}\right )  \nn \\
&&+ {1 \over 2} \sum_n {|<D^{*(n)}\left ( {1 \over 2}^-, {3 \over 2}^-\right ) (v, \varepsilon )| \vec{A}|B(v)>|^2 \over 4m_{D^{*(n)}} m_B} \ .
\eea
\noi The correction $\delta_{1/m^2}^{(A_1)}$ is therefore negative, both terms
being of the same sign.\par

The matrix elements $<D^{*(n)}\left ( {1 \over 2}^-, {3 \over
2}^-\right ) (v, \varepsilon )| \vec{A}|B>$ have been expressed in
terms of the matrix elements  $<D^{*(n)}\left ( {1 \over 2}^-\right )
(v, \varepsilon )|O_{kin, v}^{(c)}(0)|D^*(v, \varepsilon )>$ and\par
\noi $<D^{*(n)}\left ( {1 \over 2}^-, {3 \over 2}^-\right )(v, \varepsilon
)|O_{mag, v}^{(c)}(0)|D^*(v, \varepsilon )>$ by Leibovich et al. (formulas (4.1) and (4.3)) \cite{16r}. Hence, $-\delta_{1/m^2}^{(A_1)}$ (\ref{36e}) can be written as 
\bea
\label{39e}
&&- \delta_{1/m^2}^{(A_1)} = {\mu_G^2 \over 6 m_c^2} + {1 \over 8} \left ( {1 \over m_c^2} + {1 \over m_b^2} + {2 \over 3m_cm_b}\right ) \left ( \mu_{\pi}^2 - \mu_G^2\right ) \nn\\
&&+ {1 \over 2} \sum_n \left [ \left ( {1 \over 2m_c} + {3 \over 2m_b}\right ) {1 \over \Delta E_{1/2}^{(n)}}\right .  {<D^{*(n)}\left ( {1 \over 2}^-\right ) (v, \varepsilon )| O_{mag, v}^{(c)}(0)|D^*(v, \varepsilon)> \over \sqrt{4m_{D^{*(n)}} m_{D^*}}}  \nn\\
&&+ \left ( {1 \over 2m_c} - {1 \over 2m_b}\right ) {1 \over \Delta E_{1/2}^{(n)}} \left . {<D^{*(n)}\left ( {1 \over 2}^-\right ) (v, \varepsilon )| O_{kin, v}^{(c)}(0)|D^*(v, \varepsilon)> \over \sqrt{4m_{D^{*(n)}} m_{D^*}}} \right ]^2 \nn\\
&&+ {1 \over 2} \sum_n \left [  {1 \over 2m_c} \ {1 \over \Delta E_{3/2}^{(n)}}  {<D^{*(n)}\left ( {3 \over 2}^-\right ) (v, \varepsilon )| O_{mag, v}^{(c)}(0)|D^*(v, \varepsilon)> \over \sqrt{4m_{D^{*(n)}_{3/2}} m_{D^*}}} \right ]^2\ .
\eea
The important point to emphasize here is that the matrix elements\par \noi  $<D^{*(n)}\left ( {1 \over 2}^-\right ) (v,
\varepsilon )| O_{kin, v}^{(c)}(0)|D^*(v, \varepsilon)>$ and
$<D^{*(n)}\left ( {1 \over 2}^-, {3 \over 2}^-\right )(v, \varepsilon
)| O_{mag, v}^{(c)}(0)|D^*(v, \varepsilon)>$ are precisely the same ones
that enter in the SR (\ref{26e})-(\ref{28e}). This allows to obtain an
interesting lower bound on $-\delta_{1/m^2}^{(A_1)}$.

Taking now the relevant linear combinations of the matrix elements suggested by the r.h.s. of (\ref{39e}), using (\ref{26e}), (\ref{27e}) and (\ref{31e}), and Schwarz inequality
\beq
\label{42e}
\left | \sum_n A_n B_n\right | \leq \sqrt{\left ( \sum_n |A_n|^2\right ) \left ( \sum_n |B_n|^2\right )}
\eeq
 
\noi one finds
\bea
\label{43e}
&&\sum_{n\not=0} \left [ \xi^{(n)} (w)\right ]^2 \sum_{n\not= 0} \left \{ {1 \over \Delta E_{1/2}^{(n)}} \left [ \left ( {1 \over 2m_c} - {1 \over 2m_b}\right ) \right . \right . {<D^{*(n)}\left ( {1 \over 2}^-\right ) (v, \varepsilon )| O_{kin, v}^{(c)}(0)|D^*(v, \varepsilon)> \over \sqrt{4m_{D^{*(n)}} m_{D^*}}}\nn\\
&&+ \left ( {1 \over 2m_c} + {3 \over 2m_b}\right ) \left . \left . {<D^{*(n)}\left ( {1 \over 2}^-\right ) (v, \varepsilon )| O_{mag, v}^{(c)}(0)|D^*(v, \varepsilon)> \over \sqrt{4m_{D^{*(n)}} m_{D^*}}} \right ] \right \}^2 \nn \\
&& \geq 4\left \{ \left ( {1 \over 2m_c} - {1 \over 2m_b}\right )\chi_1(w) - {1 \over 3} \left ( {1 \over 2m_c} + {3 \over 2m_b}\right )\right . \left .  \left [ - 2(w-1) \chi_2 (w)+ 6 \chi_3 (w) \right ] \right \}^2 \\
&&\nn \\
&&\sum_n \left [ \tau_{3/2}^{(2)(n)}(w)\right ]^2 \sum_n \left \{ {1 \over \Delta E_{3/2}^{(n)}} \left [ {1 \over 2m_c} {<D^{*(n)}\left ( {3 \over 2}^-\right ) (v, \varepsilon )| O_{mag, v}^{(c)}(0)|D^*(v, \varepsilon)> \over \sqrt{4m_{D^{*(n)}_{3/2}} m_{D^*}}}\right ] \right \}^2\nn\\ 
&&\geq {32 \over 3} \left [ {1 \over 2m_c} \ \chi_2(w) \right ]^2\ .
\label{44e}
\eea

These two last equations imply, from (\ref{39e}), the inequality 
$$- \delta_{1/m^2}^{(A_1)} \geq {\mu_G^2 \over 6 m_c^2} + {\mu_{\pi}^2 - \mu_G^2 \over 8} \left ( {1 \over m_c^2} + {1 \over m_b^2} + {2 \over 3m_cm_b}\right )$$
$$+ \ 2 {\left \{ \left ( {1 \over 2m_c} - {1 \over 2m_b}\right ) \chi_1(w) - {1 \over 3} \left ( {1 \over 2m_c} + {3 \over 2m_b}\right ) \left [-2(w-1)\chi_2(w) + 6\chi_3 (w) \right ] \right \}^2 \over \sum\limits_{n\not= 0} \left [ \xi^{(n)}(w)\right ]^2}$$
\beq
\label{45e}
 + {16 \over 3} {\left [ {1 \over 2m_c}  \chi_2(w)\right ]^2 \over \sum\limits_n \left [ \tau_{3/2}^{(2)(n)}(w)\right ]^2}\ .
\eeq

This inequality on $-\delta_{1/m^2}^{(A_1)}$ involves on the r.h.s. {\it elastic subleading} functions $\chi_i(w)$ $(i = 1,2,3)$ in the
numerator and sums over {\it inelastic leading IW functions}
$\sum\limits_{n\not= 0} [\xi^{(n)}(w)]^2$ and $\sum\limits_n
[\tau_{3/2}^{(2)(n)}(w)]^2$ in the denominator. We must emphasize that
this inequality is valid for all values of $w$ and constitutes a
rigorous constraint between these functions and the correction
$-\delta_{1/m^2}^{(A_1)}$. Let us point out that, near $w=1$, since 
$\xi^{(n)}(w) \sim (w-1)$ $(n \not= 0)$ and, due to Luke theorem $\chi_1(w), \ \chi_3(w) \sim (w-1)$, the second term on the r.h.s. of (\ref{45e}) is a constant in the limit $w \to 1$. \par

On the other hand, since $\chi_2(w)$ is not protected by Luke theorem, $\chi_2(1) \not= 0$ and in general, as pointed out by Leibovich et al. 
$\tau_{3/2}^{(2)}(1) \not= 0$, the last term in the r.h.s. of (\ref{45e}) is also a constant for
$w = 1$. \par

The inequality (\ref{45e}) is valid for all values of $w$, and in
particular it holds in the $w \to 1$ limit. Let us consider this limit,
that gives
\bea
\label{50e}
&&- \delta_{1/m^2}^{(A_1)} \geq {\mu_G^2 \over 6 m_c^2} + {\mu_{\pi}^2 - \mu_G^2 \over 8} \left ( {1 \over m_c^2} + {1 \over m_b^2} + {2 \over 3m_cm_b}\right )\nn \\
&&+2 {\left \{ \left ( {1 \over 2m_c} - {1 \over 2m_b}\right ) \chi '_1(1) - {1 \over 3} \left ( {1 \over 2m_c} + {3 \over 2m_b}\right ) \left [-2\chi_2(1) + 6\chi '_3 (1) \right ] \right \}^2 \over \sum\limits_{n\not= 0} \left [ \xi ^{(n)'}(1)\right ]^2} \nn\\
&&+ {16 \over 3} {\left [ {1 \over 2m_c}  \chi_2(1)\right ]^2 \over \sum\limits_n \left [ \tau_{3/2}^{(2)}(1)\right ]^2}\ .
\eea

\noi On the other hand, using the OPE in the heavy quark limit, we have demonstrated above the following sum rules \cite{5newr}
\bea
\label{51e}
&&\sum_n \left [ \tau_{3/2}^{(2)}(1)\right ]^2 = {4 \over 5} \sigma^2 - \rho^2 \\
&&\sum_{n\not= 0} \left [ \xi^{(n)'}(1)\right ]^2 = {5 \over 3} \sigma^2 - {4 \over 3 }\rho^2 - (\rho^2)^2 
\label{52e}
\eea
\noi where $\rho^2$ and $\sigma^2$ are the slope and the curvature of the elastic Isgur-Wise function $\xi (w)$.

The positivity of the l.h.s. of (\ref{51e}), (\ref{52e}) yield respectively the lower bounds on the curvature obtained (\ref{22Qe}) and (\ref{27Qe}). Relations (\ref{50e})-(\ref{52e}) give finally the bound 
$$- \delta_{1/m^2}^{(A_1)} \geq {\mu_G^2 \over 6 m_c^2} + {\mu_{\pi}^2 - \mu_G^2 \over 8} \left ( {1 \over m_c^2} + {1 \over m_b^2} + {2 \over 3m_cm_b}\right ) $$
$$+ {2 \over 3[5 \sigma^2 - 4 \rho^2 - 3(\rho^2)^2]} \left \{ \left ( {1 \over 2m_c} - {1 \over 2m_b}\right ) 3\chi '_1(1) -  \left ( {1 \over 2m_c} + {3 \over 2m_b}\right ) \left [-2\chi_2(1) + 6\chi '_3 (1) \right ] \right \}^2$$ 
\beq
\label{55e}
+ {80 \over 3(4\sigma^2 - 5\rho^2)} \left [ {1 \over 2m_c} \chi_2(1)\right ]^2  \ .
\eeq

A number of remarks are worth to be made here : \par

(i) The bounds contain an OPE piece, dependent on $\mu_{\pi}^2$ and
$\mu_G^2$, and a piece that bounds the inelastic contributions, given
in terms of the $1/m_Q$ elastic quantities $\chi '_1(1)$, $\chi_2(1)$,
$\chi '_3(1)$ and of the slope $\rho^2$ and curvature $\sigma^2$ of the
elastic IW function $\xi (w)$.\par

(ii) Taking roughly constant values for $\chi '_1(1)$, $\chi_2(1)$, 
$\chi '_3(1)$, independent of $\rho^2$ and $\sigma^2$, as suggested by the QCD Sum Rules calculations (QCDSR)
\cite{21r} \cite{22r} \cite{23r}, the bounds for the inelastic
contributions diverge in the limit $\rho^2 \to {3 \over 4}$, $\sigma^2 \to {15 \over 16}$, according to (\ref{22Qe}). This
feature does not seem to us physical.\par 

(iii) {\it Therefore, one should expect that $\chi '_1(1)$, $\chi_2(1)$ and $\chi '_3(1)$ vanish also in this limit}. We give a demonstration of this interesting feature below.\par
 
(iv) Thus, the limit $\rho^2 \to {3 \over 4}$, $\sigma^2 \to {15 \over 16}$ seems related to the behaviour of $\chi_i(w)$ ($i = 1, 2, 3$) near zero recoil. \par

(v) The feature (iii) does not appear explicitly  in the QCDSR approach, where one gets roughly $\rho_{ren}^2 \cong
0.7$, and where there is no dependence on $\rho^2$ of the functions
$\chi_i(w)$ ($i = 1,2,3$).\par

Now we demonstrate that indeed $\chi '_1(1)$, $\chi_2(1)$
and $\chi '_3(1)$ vanish in the limit $\rho^2 \to {3 \over 4}$, $\sigma^2 \to {15 \over 16}$.
\noi At zero recoil $w \to 1$ we have
\beq
\label{equation73}
\chi '_1(1) = {1 \over 2} \sum_{n\not= 0} {1 \over \Delta E_{1/2}^{(n)}} \xi^{(n)'}(1) {<D^{(n)} (v) |O_{kin}^{(c)} (0)|D(v)> \over \sqrt{4m_{D^{(n)}}m_D}}
\eeq
\beq
\label{equation74}
\chi_2(1) = {1 \over 4\sqrt{6}} \sum_{n} {1 \over \Delta E_{3/2}^{(n)}} \tau_{3/2}^{(2)(n)}(1)
 {<D^{*(n)}_{3/2} (v, \varepsilon ) |O_{mag}^{(c)} (0)|D^*(v, \varepsilon )> \over \sqrt{4m_{D^{*(n)}_{3/2}}m_{D^*}}}\eeq
\beq
\label{equation75}
-4 \chi_2(1) + 12\chi '_3(1) = \sum_{n\not= 0} {1 \over \Delta E_{1/2}^{(n)}} \xi^{(n)'}(1)
 {<D^{(n)} (v) |O_{mag}^{(c)} (0)|D(v)> \over \sqrt{4m_{D^{(n)}}m_D}}
\eeq

\noi Using again Schwarz inequality, we obtain 
\beq
\label{equation76}
[\chi '_1(1)]^2 \leq  {1 \over 4} \sum_{n\not= 0} \left [\xi^{(n)'}(1)\right ]^2 \sum_{n\not= 0} \left [ {1 \over \Delta E_{1/2}^{(n)}}  {<D^{(n)} (v) |O_{kin}^{(c)} (0)|D(v)> \over \sqrt{4m_{D^{(n)}}m_D}}\right ]^2
\eeq
\beq
\label{equation77}
[\chi_2(1)]^2 \leq  {1 \over 96} \sum_{n} \left [ \tau_{3/2}^{(2)(n)}(1)\right ]^2\sum_{n} \left [ {1 \over \Delta E_{3/2}^{(n)}}  {<D^{*(n)}_{3/2} (v, \varepsilon) |O_{mag}^{(c)} (0)|D^*(v, \varepsilon)> \over \sqrt{4m_{D^{(n)}}m_{D^*}}}\right ]^2
\eeq
\beq
\left [ -4 \chi_2(1) + 12\chi '_3(1)\right  ]^2 \leq  \sum_{n\not= 0}  \left [ \xi^{(n)'}(1) \right ]^2
 \sum_{n\not= 0} \left [ {1 \over \Delta E_{1/2}^{(n)}}{<D^{(n)} (v) |O_{mag}^{(c)} (0)|D(v)> \over \sqrt{4m_{D^{(n)}}m_D}}\right ]^2
\label{equation78}
\eeq
\noi and from relations (\ref{51e}) and (\ref{52e}),
\beq
\label{equation79}
[\chi '_1(1)]^2 \leq  {1 \over 12}  \left [5 \sigma^2 - 4 \rho^2 - 3(\rho^2)^2\right ] \sum_{n\not= 0} \left [ {1 \over \Delta E_{1/2}^{(n)}}  {<D^{(n)} (v) |O_{kin}^{(c)} (0)|D(v)> \over \sqrt{4m_{D^{(n)}}m_D}}\right ]^2\eeq
\beq
\label{equation80}
[\chi_2(1)]^2 \leq {1 \over 480}  \left (4 \sigma^2 - 5 \rho^2) \right )\sum_{n} \left [ {1 \over \Delta E_{3/2}^{(n)}}  {<D^{*(n)}_{3/2} (v, \varepsilon ) |O_{mag}^{(c)} (0)|D^*(v, \varepsilon )> \over \sqrt{4m_{D^{*(n)}_{3/2}}m_{D^*}}}\right ]^2\eeq
\bea
&&\left [ -4 \chi_2(1) + 12\chi '_3(1)\right  ]^2\nn \\
&& \leq  {1 \over 3}  \left [ 5 \sigma^2 - 4 \rho^2 - 3(\rho^2)^2\right ] 
\sum_{n\not= 0}\left [ {1 \over \Delta E_{1/2}^{(n)}}{<D^{(n)} (v) |O_{mag}^{(c)} (0)|D(v)> \over \sqrt{4m_{D^{(n)}}m_D}}\right ]^2
\label{equation81}
\eea
\noi Therefore, in the limit $\rho^2 \to {3 \over 4}$, $\sigma^2 \to {15 \over 16}$, one obtains
\beq
\label{equation82}
\chi ' _1(1) = \chi_2(1) =  \chi '_3(1) = 0
\eeq

This is a very strong correlation relating the behaviour of the elastic IW function $\xi (w)$ to the elastic subleading IW functions $\chi_i (w)$ ($i = 1,2,3$) near zero recoil.

To conclude, we have obtained bounds that relate the $1/m_Q^2$ correction
of the form factor $h_{A_1}(w)$ to the $1/m_Q$ subleading form factors of the
Lagrangian type $\chi_i(w)$ ($i=1,2,3$) and to the shape of the elastic
Isgur-Wise $\xi (w)$. This bound should in principle be taken into
account in the analysis of the exclusive determination of $|V_{cb}|$ in
the channels $\overline{B} \to D(D^*)\ell \nu$. On the other hand, we have demonstrated an important constraint on the behavior of the subleading form factors $\chi_i(w)$~: in the limit
$\rho^2 \to {3 \over 4}$, $\sigma^2 \to {15 \over 16}$, $\chi'_1(1)$, $\chi_2(1)$ and $\chi '_3(1)$ must vanish.\par

It would be very interesting to have a theoretical calculation of the functions $\chi_i(w)$ ($i = 1,2,3$) satisfying this constraint. Otherwise it seems questionable to try an exclusive determination of $|V_{cb}|$ by fitting the slope $\rho^2$ and considering uncorrelated subleading corrections.

In conclusion, using sum 
rules in HQET, as formulated in ref. 
\cite{3r,5newr,9r,19r}, we
have found lower bounds for the moduli of the derivatives of $\xi (w)$ and non-trivial results on $1/m_Q$ form factors of the Current and Lagrangian types. We have also obtained a lower bound on $-\delta_{1/m^2}^{(A_1)}$. The determination of the
CKM matrix element $|V_{cb}|$ in $B \to
D^{(*)}\ell \nu$ should satisfy 
these constraints.

\section*{Acknowledgement} We are indebted to the EC contract
HPRN-CT-2002-00311 (EURIDICE) for its support.

\end{document}